\documentclass[useAMS,usenatbib]{mn2e}
\usepackage{epsfig}

\title[The first rotation periods in Praesepe]{The first rotation periods in Praesepe}

\author[Alexander Scholz \& Jochen Eisl{\"o}ffel]{Alexander Scholz$^{1}$\thanks{E-mail:
as110@st-andrews.ac.uk} and Jochen Eisl{\"o}ffel$^{2}$\thanks{E-mail: jochen@tls-tautenburg.de}\\
$^{1}$SUPA, School of Physics \& Astronomy, University of St. Andrews, North Haugh, St. Andrews, Fife KY16 9SS, United Kingdom\\
$^{2}$Th{\"u}ringer Landessternwarte Tautenburg, Sternwarte 5, D-07778 Tautenburg, Germany}

\begin{document}

\date{Accepted. Received.}

\pagerange{\pageref{firstpage}--\pageref{lastpage}} \pubyear{2002}

\maketitle

\label{firstpage}

\begin{abstract}
The cluster Praesepe (age $\sim 650$\,Myr) is an ideal laboratory to study stellar evolution. Specifically, it 
allows us to trace the long-term decline of rotation and activity on the main-sequence. Here we present rotation
periods measured for five stars in Praesepe with masses of $0.1-0.5\,M_{\odot}$ -- the first rotation periods 
for members of this cluster. Photometric periodicities were found from two extensive 
monitoring campaigns, and are confirmed by multiple independent test procedures. We attribute these 
variations to magnetic spots co-rotating with the objects, thus indicating the rotation period. The five 
periods, ranging from 5 to 84\,h, show a clear positive correlation with object mass, a trend which
has been reported previously in younger clusters. When comparing with data for F-K stars in the coeval
Hyades, we find a dramatic drop in the periods at spectral type K8-M2 (corresponding to $0.4-0.6\,M_{\odot}$).
A comparison with periods of VLM stars in younger clusters provides a constraint on the spin-down
timescale: We find that the exponential rotational braking timescale is clearly longer than 200\,Myr, 
most likely 400-800\,Myr. These results are not affected by the small sample size in the rotation periods
in Praesepe. Both findings, the steep drop in the period-mass relation and the long spin-down 
timescale, indicate a substantial change in the angular momentum loss mechanism for very low mass (VLM) 
objects, possibly the breakdown of the solar-type (Skumanich) rotational braking. While the physical 
origin for this behaviour is unclear, we argue that parts of it might be explained by the disappearance 
of the radiative core and the resulting breakdown of an interface-type dynamo in the VLM regime. Rotational 
studies in this mass range hold great potential to probe magnetic properties and interior structure of 
main-sequence stars.

\end{abstract}


\section{Introduction}

Clusters are ideal environments to study the evolution of stars. Members of one particular cluster
share age, distance, as well as metallicity, and these parameters are typically well-constrained. Thus, clusters
provide 'snapshots' of the stellar population at a given age. Comparing the properties of stars in
clusters of different ages allows us then to track their evolution. Furthermore, clusters are often dense
enough to allow multiplexed observations, i.e. wide-field imaging or multi-object spectroscopy, 
covering a large sample of objects simultaneously.

Two key parameters of stars, which strongly depend on age, are rotation and magnetic activity. The fundament 
for the understanding of the main-sequence evolution of rotation and activity is the empirically found 
Skumanich law: Looking at averages of large samples of objects, the rotational velocities of F to K 
stars evolve proportional to the inverse squareroot of age \citep[e.g., ][]{1972ApJ...171..565S,2001ApJ...561.1095B}. 
An analogous decline is seen for indicators of magnetic activity, implying a correlation between rotation 
and activity, which has been confirmed in multiple studies \citep[e.g., ][]{1997ApJ...479..776S,2000AJ....119.1303T}. 
This is usually interpreted with a twofold connection between stellar rotation and magnetic activity, providing a strong 
coupling between those two parameters: a) The interior dynamo of F-K stars, i.e. the underlying mechanism to generate 
magnetic activity, is thought to be governed by rotation. b) The rotational braking on the main-sequence is due to 
angular momentum losses through magnetic stellar winds \cite[see][]{2000ssma.book.....S}.

To explain the presence of large-scale, stable magnetic fields in solar-type stars, it is usually 
assumed that the dynamo in F-K stars is similar to the solar-type $\alpha\omega$ dynamo and operates in the 
transition layer between the convective envelope and the radiative core. Therefore, it is expected that the 
magnetic field generation, and thus the magnetic and rotational properties, change when objects become fully 
convective at masses $<0.3-0.4\,M_{\odot}$ \citep{1997A&A...327.1039C}.
However, no clear change of rotation and activity indicators has yet been detected at or around this mass limit 
\citep[e.g., ][]{1998A&A...331..581D}. In addition, it is unclear what type of magnetic 
field is harboured by fully convective objects, and if it depends on rotation. Investigating the 
rotational evolution and angular momentum loss of very low mass (VLM) objects on the main-sequence
can contribute to clarify these issues.

An ideal laboratory to analyse the rotation of VLM main-sequence stars is Praesepe: With an age probably between 
0.6 and 0.9\,Gyr \citep[see][]{1981A&A....97..235M,2002AJ....124.1570A}, the cluster is significantly older than zero age 
main sequence (ZAMS) clusters like the Pleiades or $\alpha$\,Per. At these ages, all stars have safely arrived on the 
main-sequence and the rotational regulation is thus dominated by magnetic wind braking, and not anymore affected by initial 
conditions, disk-braking, and contraction -- the objects have essentially 'forgotten' their pre-main sequence history.
Thus, it is possible to isolate the effect of wind braking on the rotational properties.

Compared with the similarly old clusters Hyades and Coma, Praesepe is relatively compact, yet
not too far away \cite[180\,pc, ][]{1999A&A...345..471R}, allowing us deep, efficient observations with wide-field 
facilities. Here we report on photometric monitoring for 18 members of Praesepe, which provided the first five 
rotation periods for members of this cluster -- all for objects with estimated masses below 0.5$\,M_{\odot}$. Combined with 
previously published periods for younger objects \citep{2004A&A...419..249S,2004A&A...421..259S,2005A&A...429.1007S}, 
this allow us to probe the long-term evolution of rotation in the VLM regime.\footnote{Based on observations collected 
at the Centro Astronomico Hispano Aleman (CAHA) at Calar Alto, operated jointly by the Max-Planck Institut f{\"u}r Astronomie 
and the Instituto de Astrofisica de Andalucia (CSIC), and at the Th{\"u}ringer  Landessternwarte Tautenburg (Germany)}

\section{Photometric monitoring}

\subsection{Observations}

The analysis in this paper is based on two photometric time series of VLM stars in Praesepe. The observing 
strategy in both runs was similar: To mimimize internal inconsistencies in the data and thus systematic uncertainties,
we 'stared' at a particular field in Praesepe. We did not 'dither' around a central position and made an effort 
to position the field of view at the same coordinates in each observing night. Therefore each object was at 
roughly the same pixel position on the detector throughout the run. The exposure time for the single frames was 
600\,sec in both runs. Since we were interested in very low mass objects, which have a spectral energy distribution 
peaking in the near-infrared, all observations were carried out in the I-band filter.

The target fields were selected to maximise the number of known Praesepe members in the field of view. We compiled
an initial catalogue of VLM members from the surveys of 
\citet{1995MNRAS.273..505H,1995A&AS..109...29H,1997MNRAS.287..180P,1998ApJ...497L..47M}, at the time of the observations
(2001, 2003) essentially the complete census of the VLM population in this cluster. For all these objects, the main
criterium for cluster membership is multi-colour photometry, confirmed in many cases by proper motions and/or
spectroscopy. According to a follow-up study by \citet{1999MNRAS.310...87H}, the contamination rate is $\sim 10$\% 
in the Hambly et al. sample and $\sim 50$\% among the (fainter) Pinfield et al. objects. 

The first time series was obtained using the 2-m Schmidt telescope at the Th{\"u}ringer Landessternwarte Tautenburg 
(TLS, Germany), equipped with a $2048 \times 2048$ SiTe CCD camera. The wide-field camera provides a spatial resolution 
of 1\farcs2/pixel, resulting in an unvignetted field of view of about 0.36\,sqdeg. The observations at the TLS were 
carried out in two sessions in January and February 2001, with a gap of almost four weeks. Thus, although we cover more 
than a month in total, our sensitivity to long periods (1-4 weeks) will be limited. The observations were partly affected 
by cirrus and mediocre seeing conditions. In total, the run provided a time series of 125 images of the same field. The 
target field for the TLS run was centred at $\alpha = 8^h36^m53^s$, $\delta = +19^\circ 48\arcmin 24\arcsec$ (J2000.0). 
This field contains nine members from the Hambly et al. catalogue (H85, 91, 102, 106, 115, 126, 140, 158, 181 in their 
nomenclature) and two from the Pinfield survey (P1 and P2 in their nomenclature), plus some more higher-mass members 
which are saturated in our images.

\begin{table}
\centering
\caption{Logfile of the time series observations: Run, date of observations, no. of images, typical seeing}
\begin{tabular}{llcc}
\hline
Run & Date & No.& Seeing\\
\hline
TLS & 16/01/2001 & 12 & 2\farcs0\\
    & 17/01/2001 & 12 & 2\farcs5\\
    & 18/01/2001 & 6  & 2\farcs0\\
    & 14/02/2001 & 32 & 2\farcs0\\
    & 15/02/2001 & 38 & 1\farcs8\\
    & 16/02/2001 & 24 & 1\farcs8\\
    & 18/02/2001 & 1  & 2\farcs5\\
\hline
LAICA & 23/01/2003 & 13 & 3\farcs8\\
      & 24/01/2003 & 19 & 2\farcs5\\
      & 25/01/2003 & 11 & 2\farcs0\\
      & 27/01/2003 & 22 & 1\farcs5\\
      & 28/01/2003 & 43 & 1\farcs3\\
\hline
\end{tabular}
\end{table}

A second campaign was carried out using the wide-field imager LAICA at the 3.5-m telescope at Calar Alto Observatory 
from 23-28 Jan 2003 (in service mode). LAICA is a $2 \times 2$ mosaic of $4\times 4$K CCDs, which we used to 'stare' at 
our target field in Praesepe. Thus, we did not observe a continuous field, but four 'patches' of $15'\times 15'$, separated 
by gaps of similar size, in total 0.25\,sqdeg at a resolution of 0\farcs22/pixel. The total length of the campaign was 10 
nights (scheduled as 10 half nights), but most of the first half of the run was not useful due to excessively high seeing and 
cloud coverage. In the last six nights, however, we were able to obtain in total 108 images with mostly good quality. 
Thus, this time series will be highly sensitive to periods up to a few days. The LAICA run was focused on a target field 
at $\alpha = 8^h40^m56.5^s$, $\delta = +19^\circ 32\arcmin 30\arcsec$ (J2000.0), not overlapping with the TLS field. 
Since the LAICA images are considerably deeper than the TLS data, we aimed to cover some of the faintest member 
candidates in the cluster. The field thus contains six objects from the Pinfield survey (P14, 15, 16, 17, 19, 20)
plus one additional unsaturated star from the Hambly list (H218). 

\subsection{From images to lightcurves}

Image reduction, photometry, and relative calibration for both runs followed the routines established in the framework of
our previous campaigns \citep{2004A&A...419..249S,2004A&A...421..259S,2005A&A...429.1007S}. In the following,
we give a brief account of the basic principles of our method to derive lightcurves from time series images. For
details, we refer to the more elaborated discussion in the afore mentioned papers, particularly \citet{2004A&A...419..249S}.
In case of the LAICA campaign, all reduction and calibration steps were carried out separately on the four individual
chips.

The image reduction started with bias subtraction and flatfield correction (based on high signal-to-noise domeflats).
To remove the interference ('fringe') pattern produced by night sky line emission, which is typical for deep I-band
exposures, we constructed a fringe mask (an image of the fringes, without objects) by averaging a number of dark sky
exposures. This fringe mask was then properly scaled and subtracted from the individual time series images. Scaling is
necessary because the amplitude of the fringes is a function of airmass and observing conditions. After this process, 
the residuals of the fringes in the final images do not exceed 1\% of the sky background.

Using the Source Extractor, we created an object catalogue for one selected image in each time series ('master image').
Customised IRAF routines were used to a) determine pixel offsets between the master frame and all individual time
series images, b) creating object catalogues for all time series images, c) carry out aperture photometry for 
all objects. For the TLS run, PSF fitting based on {\it daophot} was chosen as the optimum method for photometry. 
We selected a sample of PSF reference stars and used them to model the PSF in all time series images. Based
on the values from the aperture photometry, new magnitudes were derived by adapting the model PSF to all objects
in the field.

We refrained from using PSF fitting techniques for the LAICA run: Due to the strongly variable seeing conditions,
it was not possible to find a set of PSF reference stars usable for a complete time series. Changing the reference 
PSF, however, may introduce systematic uncertainties. On the other hand, LAICA has an excellent pixel resolution, 
diminishing the benefits of PSF fitting photometry in moderately crowded fields. Therefore, we rely on the more 
robust aperture technique.

The instrumental magnitudes are still affected by the combined effects of varying airmass, varying atmospheric extinction,
and variable zeropoint. This is corrected by subtracting a 'master lightcurve' from all time series, which is an average
lightcurve calculated from a set of non-variable reference stars. The selection of these reference stars is the 
critical part in the relative calibration of the photometry. In a first step, we chose only objects with valid photometry
and photometric error below 0.1\,mag in all images. From this initial set (typically a few hundred objects), we excluded 
potentially variable objects by testing the rms in their lightcurve against the rms of all other preliminary reference 
stars. In several steps, the sample of reference stars is cleared from contaminating variable objects. For details 
of the process, we refer to \citet{2004A&A...419..249S}. After subtracting the resulting master lightcurve from all time 
series, we obtained lightcurves in relative magnitudes for all objects in our fields. This database is the fundament for 
the variability analysis described in Sect. \ref{perser}.

We give two estimates for the accuracy of the relative calibration: a) The average rms of the final lightcurves of the 
reference stars (found to be non-variable) is 0.01\,mag for the TLS and 0.02\,mag for the LAICA run. b) The
minimum rms measured from the lightcurves of bright stars is 0.009\,mag in the TLS run and 0.01\,mag for the LAICA run.
Both methods provide an estimate of the intrinsic noise in the lightcurves. We conclude that for bright, but unsaturated
objects we achieve a photometric accuracy of about 1\% in both runs.

\section{Period search}
\label{perser}

The main goal of the photometric monitoring was to detect periodic changes in the flux of our targets, which can be 
interpreted as indication of co-rotating spots -- thus to measure rotation periods. The lightcurves of the Praesepe
members in our fields were scrutinised with a rigorous period search procedure, which has been implemented and extensively
tested in the framework of previous campaigns \citep{2004A&A...419..249S,2004A&A...421..259S,2005A&A...429.1007S}. 

\begin{table}
\caption{Objects with significant periodic variability in Praesepe: Object ID
(following \citet{1995A&AS..109...29H,1997MNRAS.287..180P}), mass estimate (see Sect. \ref{rotmass}), 
spectral type estimate (see Sect.\ref{rotmass}), period, period uncertainty, period amplitude}
\label{periods}
\begin{tabular}{lcccccc} 
\hline
Object	 & M ($M_{\odot}$) & Sp.T. & P (h) & $\Delta$P (h) & A (mag)\\
\hline
{\it TLS:}\\
P2   & 0.12 & M6.5 & 5.65 & 0.01 & 0.04 \\
H91  & 0.29 & M3 & 42.2 & 0.20 & 0.08 \\
H115 & 0.40 & M2 & 83.6 & 1.02 & 0.04 \\
{\it LAICA:}\\
P20  & 0.11 & M7 & 12.2 & 0.20 & 0.07 \\
H218 & 0.19 & M4 & 16.2 & 0.51 & 0.04 \\
\hline             
\end{tabular}			    
\end{table}	      

Period search in astronomic time series is non-trivial, mainly because of the characteristic gaps caused
be daytime, weather changes, and unavailibility of the telescope. As a result, the distribution of datapoints is
usually strongly clumped, hampering time series analysis. This is further complicated in the case of very low mass
objects, where signal-to-noise (defined as ratio between amplitude of the periodic variation and noise level in the
lightcurve) is typically less than five, so that the period is not always obvious from a by-eye examination. We try
to mitigate these problems by a) maximising the number of datapoints in the lightcurve (to obtain as much information
as possible about the light changes), and b) by probing for periodicities using a variety of independent and robust
tests. In the following, we briefly outline the criteria which we require to be fulfilled to accept a period:

\begin{enumerate}
\item{The Scargle periodogram \citep{1982ApJ...263..835S} shows a peak with a preliminarily determined false alarm probability 
(FAP) below 1\% (FAP calculated following \citet{1986ApJ...302..757H}).}
\item{Nearby reference stars do not show the same peak in the periodogram.}
\item{The scatter in the lightcurve is significantly (based on a F-test) reduced after subtraction of the sine-wave
approximated period.}
\item{The lightcurves of reference stars phased to the detected period do not show any sign of periodicity.}
\item{The CLEAN algorithm \citep{1987AJ.....93..968R} confirms the periodogram peak, i.e. it is not an artefact 
from the window function.}
\item{The empirical FAP determined using a bootstrap approach (see below) is below 1\%.}
\item{The phaseplot shows the period clearly (but see below).}
\end{enumerate}

\begin{figure}
\center
\includegraphics[width=4.7cm,angle=-90]{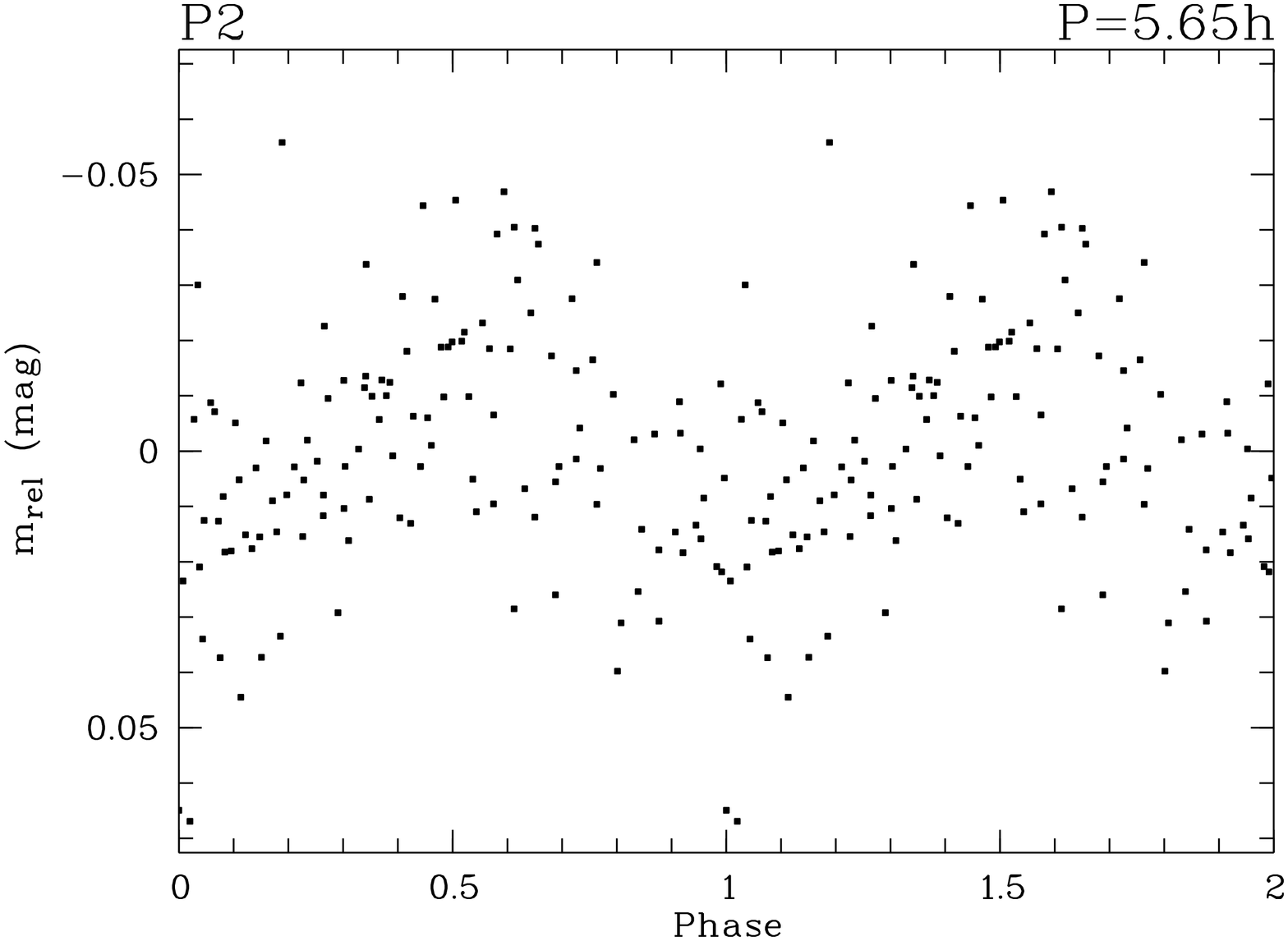} 
\includegraphics[width=4.7cm,angle=-90]{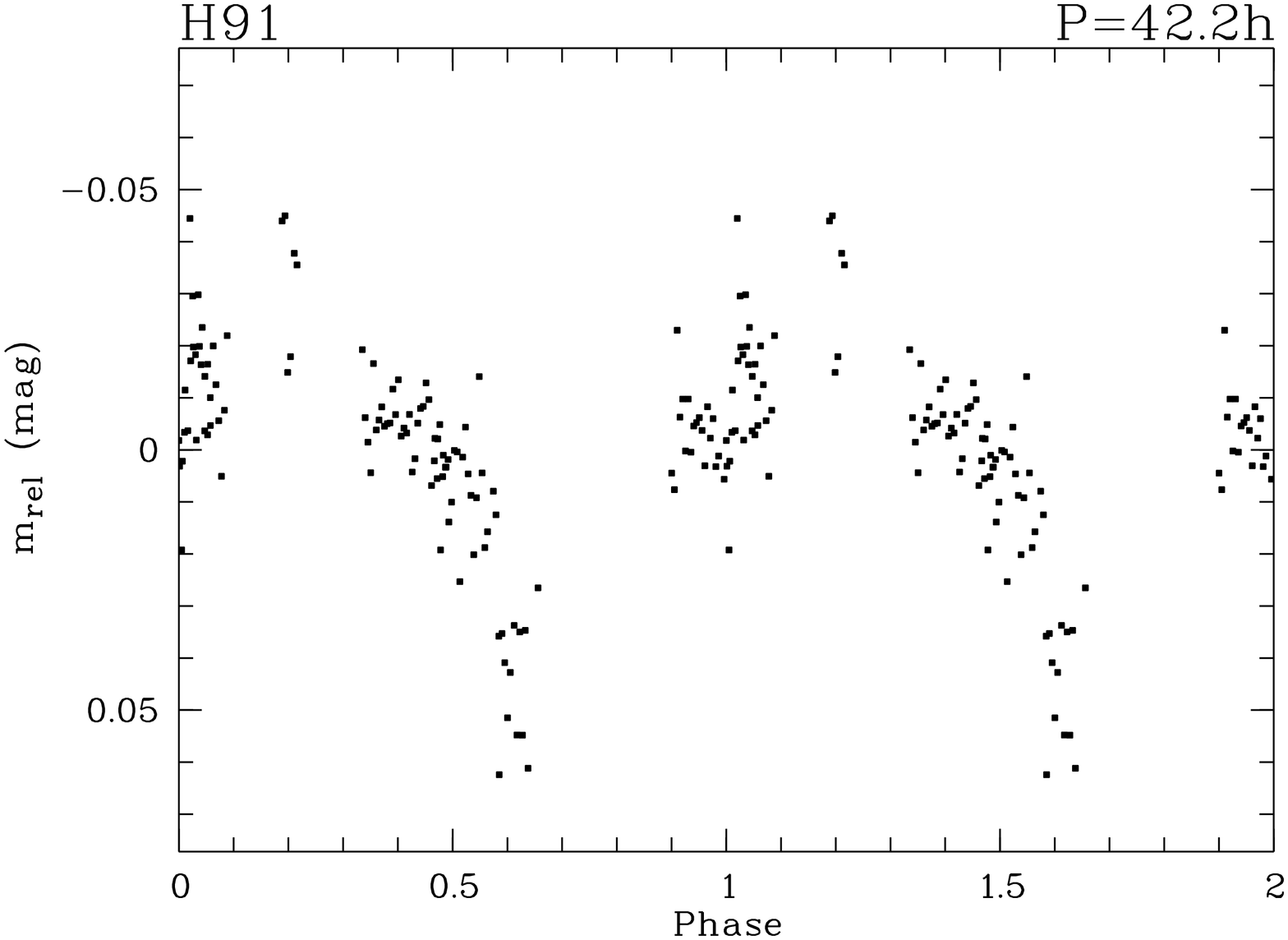} 
\includegraphics[width=4.7cm,angle=-90]{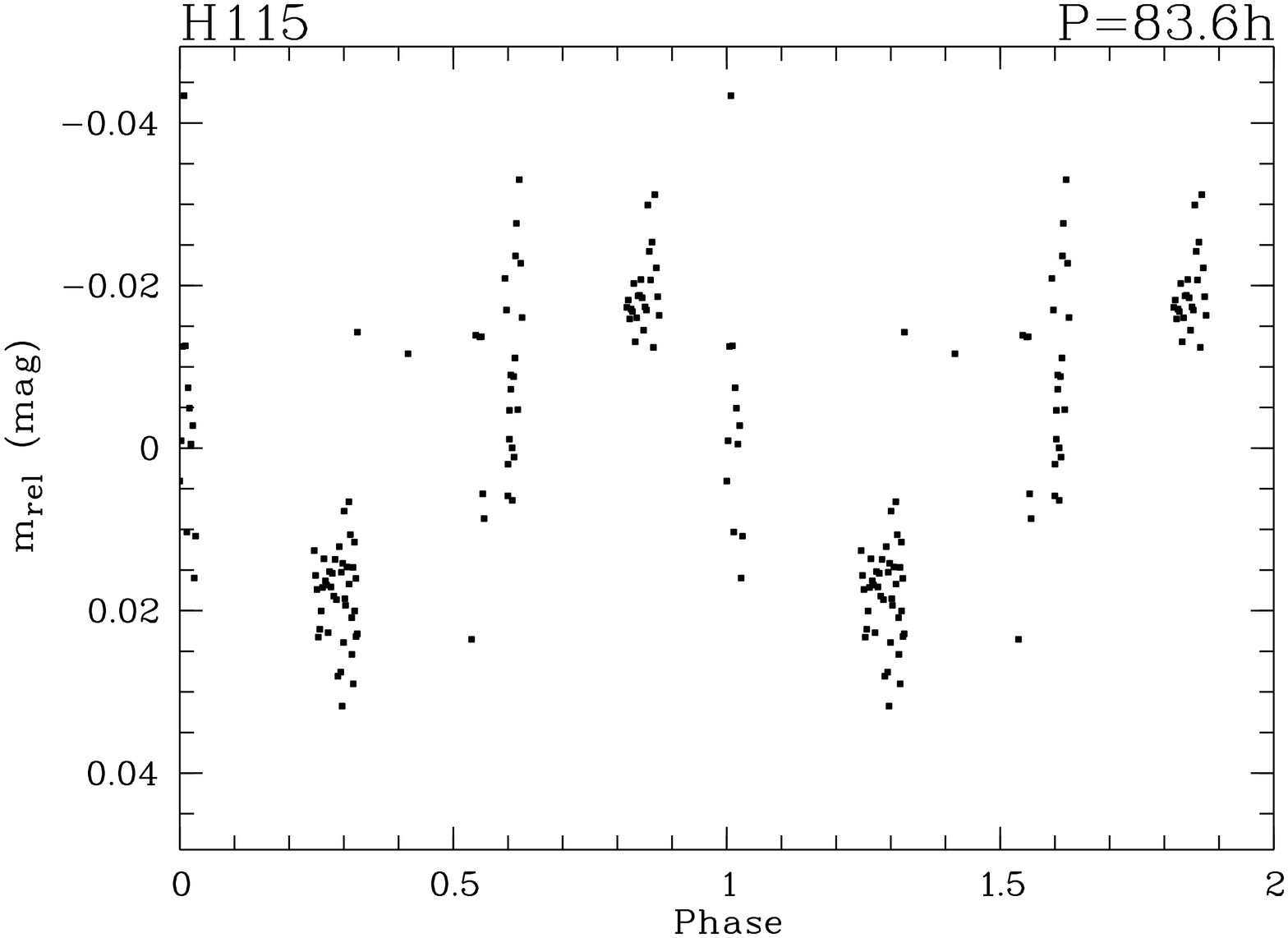} 
\includegraphics[width=4.7cm,angle=-90]{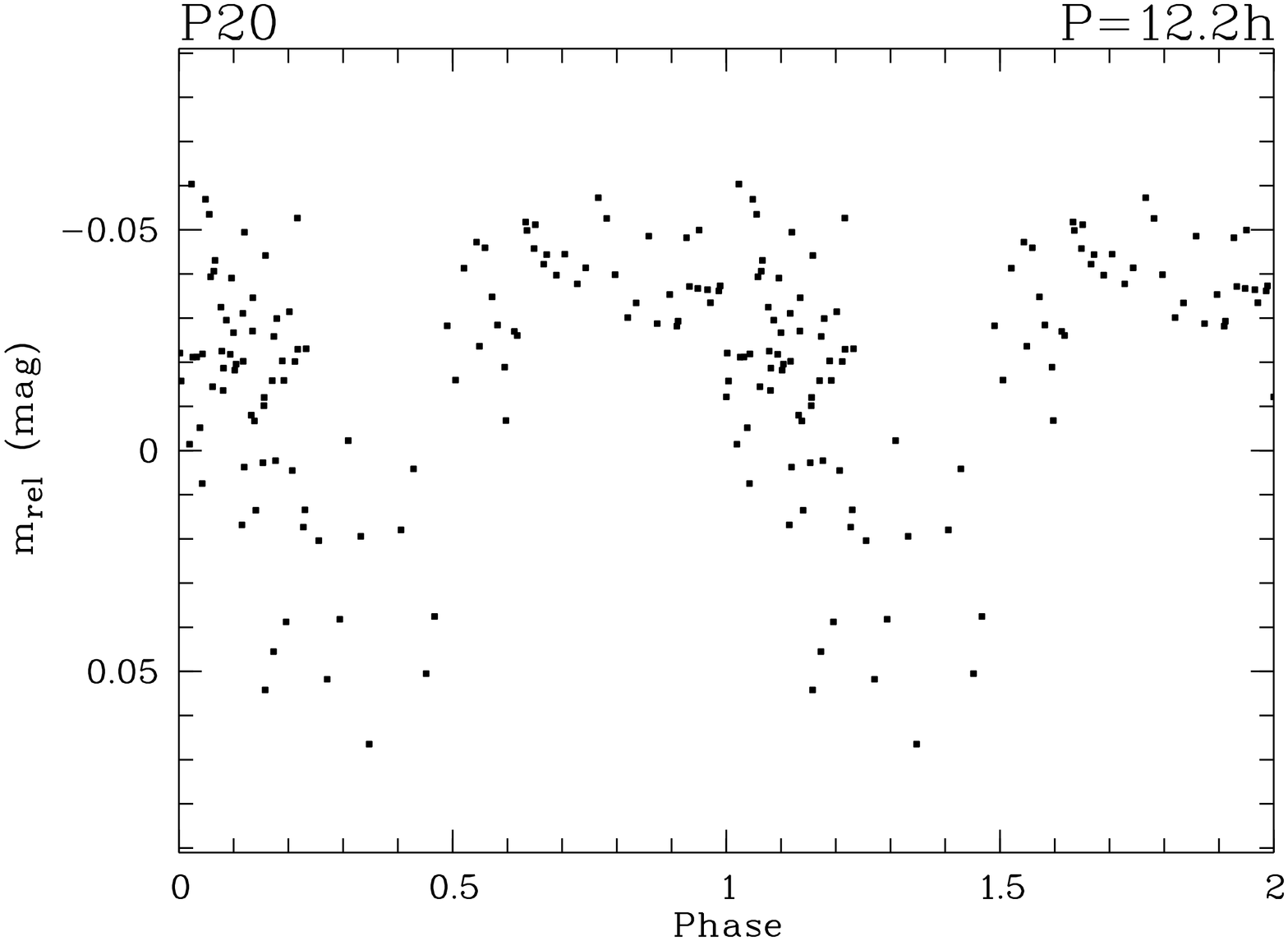} 
\includegraphics[width=4.7cm,angle=-90]{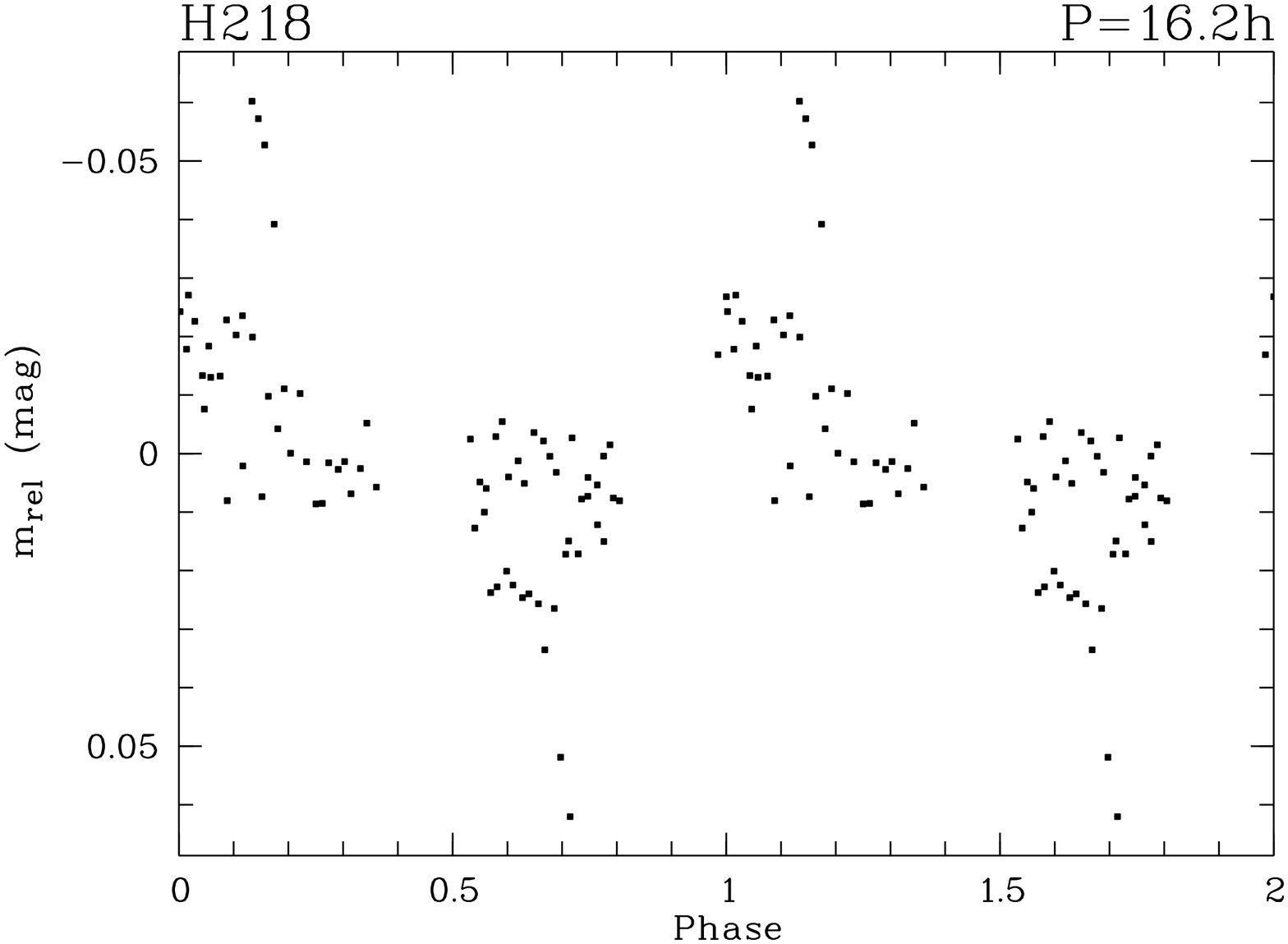} 
\caption{Phase plots for the five objects with significant periodic variations. P2, H91, H115 have been
observed in the TLS run, P20 and H218 in the LAICA run \label{f1}.}
\end{figure}

Applying these criteria, we identified five objects with significant photometric period in the lightcurve
(see Table \ref{periods}). The phaseplots for these five objects are shown in Fig. \ref{f1}. Please see below for 
individual notes on the lightcurves. Final values for the FAP have been determined using 
the bootstrap approach explained in detail in \cite{2004A&A...419..249S}. The basic idea is to produce 10000 randomised 
lightcurves per object by retaining the observing times and shuffling the data-values only. The resulting lightcurves 
will have the same sampling and noise level as the original time series, but certainly no periodicity. For all 10000 
random lightcurves we calculated the Scargle periodogram and determined the amplitude of the highest peak. The fraction 
of lightcurves, for which this number exceeds the peak amplitude for the {\it measured} periodicity of the given object 
is a robust and reliable estimate for the FAP for this period. For all five detected periods, the peak in the
periodogram is clearly higher than any peak from the 10000 randomised lightcurves; thus the FAP is below 0.01\%. 

We tested the range of periods for which we are sensitive by adding sinewaves with typical amplitudes of 0.05\,mag and 
varying period to the lightcurves of non-variable reference stars. By applying the period search procedure used for our 
target stars to these 'artificial' periodic objects, we find that our sampling allows us to reliably detect periods between 
$\sim 1$\,h and $\sim 4$\,days (in both runs). This is in line with the expectations: Both monitoring campaigns provided dense 
sampling over at least three consecutive nights, with only spotty coverage for longer timescales. Hence, it is not surprising 
that all our five detected periods are in the range between 1-4\,d. However, it has to be emphasised that the period search 
is not sensitive for periods around multiples of 1\,day, due to the daytime gaps, thus the period sample is probably not complete. 
On the other hand, both time series have some potential to detect periods longer than 4\,days (up to periods of a month for the 
TLS run). Still, given the fact that our sensitivity for these long periods is limited, we do not put too much emphasis
on the fact that we did not find any such period.

Our period search still includes one subjective criterium: the by-eye examination of the phased lightcurves. We require
that the period is visible in this plot, i.e. that it shows a clear maximum and minimum. However, this criterium  has
to be used with caution. As we have shown in \citet{2005A&A...429.1007S}, periodogram techniques are able to pick
up periods even at signal-to-noise levels too low to produce a visually obvious periodicity in the lightcurve. Thus,
a not entirely convincing phaseplot (see Fig. \ref{f1}) does not necessarily indicate a false detection. This has
to be taken into account when discussing individual objects. Please note that the objects P2, H115, and P20 have
been confirmed as cluster members by near-infrared photometry \citep{1999MNRAS.310...87H}. Based on their position
in the colour-magnitude diagram, P2 and H115 have been classified as possible binaries.

Notes on individual objects:

{\it P2:} The object is one of the faintest, lowest mass members of Praesepe discovered so far. We estimate roughly
0.1$\,M_{\odot}$; if the object really is a binary, this corresponds to the total system mass. Additionally, the amplitude
of the periodicity if fairly low, resulting in low signal-to-noise ratio. This is a typical case where the effects of the
low signal-to-noise are offset by a large number of datapoints (in this case 123). Scargle and CLEAN periodogram 
show a convincing peak, which translates into a highly significant period of 5.65\,h. 

{\it H91:} Although the phaseplot for the period of 42.2\,h exhibits some gaps, the datapoints do show a clear flux modulation,
i.e. maximum and minimum are clearly visible.

{\it H115:} Due to the relatively long period of 83.6\,h, the coverage is not continuous. The
two strong clumps at phases of 0.3 and 0.85 represent the datapoints from Feb 14 and 16, respectively, which are clearly
offset by roughly 0.04\,mag, and thus give direct evidence for a photometric variation on timescales of days.

{\it P20:} The phaseplot shows a clear periodicity, but the lightcurve appears 'box'-shaped. This raises the question whether
we are dealing with an eclipsing binary rather than a rotational modulation. However, when examining the lightcurve,
there are clear smooth and gradual trends in some nights rather than rapid brightness changes, which is difficult to 
explain in an eclipse scenario. Moreover, the fact that the suspected eclipse covers almost half of the phase space 
would require a roughly equal-radius binary system, which is unlikely given the very small eclipse depth. Moreover,
the period would imply an orbital separation of fractions of 1\,AU, and such systems seem to be rare \citep{1992ApJ...396..178F}.
Therefore, we attribute the 'boxy' shape of the lightcurve to an unusual spot configuration. Still, the possibility of an 
eclipsing system cannot completely be disregarded.

{\it H218:} The largest peak in the Scargle periodogram corresponds to a period of 33.6\,h, but disappears after applying
the CLEAN routine and is thus probably an artefact. CLEAN, however, identifies a peak at the doubled frequency (corresponding 
to a period of 16.2\,h) as real. Since the same peak is also highly significant in the Scargle periodogram, we adopt a period 
of 16.2\,h for this object. The period remains significant when we exclude the five outlying datapoints at phases of 
0.15 and 0.7.

\section{Interpretation}

The best explanation for the observed periodic variability is the presence of spots on the surface of the 
stars, co-rotating with the objects and thus modulating the flux. With an idealised spot distribution (one spot, regularly
shaped), we would expect sine-shaped periodicities, particularly at low signal-to-noise ratio. In contrast, other sources
of periodic variations (eclipsing binaries, planetary transits) produce regular dips in the lightcurves. Four out of five 
periods clearly resemble a sinecurve. The sole exception, P20, could be an eclipsing binary, but we argue that its box-shaped
lightcurve is more likely caused by a particular distribution of surface features. 

The underlying reason of the surface features is most likely magnetic activity: Our targets have effective temperatures 
larger than 3000\,K (spectral types around mid M), rendering the possibility of condensated dust clouds, as often discussed 
for the much cooler L dwarfs, improbable. M dwarfs, on the other hand, are known to harbour substantial magnetic fields, as 
evidenced by X-ray and H$\alpha$ activity. Specifically, for three of our periodic objects (H91, H115, H218) chromospheric
activity evidenced by H$\alpha$ emission with equivalent widths between 3 and 23\,\AA~has been reported by
\citet{1998ApJ...506..347B} and \citet{2006AJ....132.1517K}. It seems reasonable to assume that the remaining two objects 
(without H$\alpha$ observation so far) share similar chromospheric properties.

Thus, for the following discussion, we assume that the most likely origin for the observed periodic variations is a rotational 
modulation due to magnetically induced surface spots. Hence, the periods correspond to the rotation periods of our targets. 
These are the first rotation periods measured for objects in this mass range in a cluster significantly older than the ZAMS 
stage (which is reached at roughly 0.2-0.3\,Gyr for very low mass stars, see \citet{2007MNRAS.tmp..276I}). In the following 
subsections, we will use this small period sample in combination with literature data to probe rotation vs. mass and rotation 
vs. age for very low mass objects on the main-sequence.

To assure ourselves that the five datapoints give a realistic first glance of the VLM period distribution in Praesepe, we carried
out two plausibility checks. First, we compare with $v\sin i$ values for the Hyades, compiled by \citet{2000AJ....119.1303T}. 
The Hyades are roughly coeval with Praesepe, and since the $v\sin i$ sample size in the very low mass range is about 20,
these values can provide a useful consistency check. In the mass range 0.2-0.4\,$M_{\odot}$, the lower limit in rotational 
velocities is defined by the detection limit of 5\,kms$^{-1}$, which translates to an upper limit in $P\sin i$ of 30-90\,h,
depending on mass (with radii from \citet{1997A&A...327.1039C}). The upper limit in $v\sin i$ is around 15\,kms$^{-1}$, 
corresponding to a lower limit for $P\sin i$ between 10 and 30\,h. Considering that $\sin i$ has a fairly tight distribution 
around 0.7-0.8 for random orientations of rotational axes, these numbers are completely consistent with our period limits.

A second check can be made based on the period ranges for clusters somewhat younger than Praesepe. In the Pleiades,
NGC2516, and M34, all clusters with ages between 100 and 200\,Myr, a substantial sample of VLM periods has been established, 
in total $\sim 300$, the dominant majority in NGC2516 \citep{2006MNRAS.370..954I,2007MNRAS.tmp..276I,2004A&A...421..259S}. These 
periods range from 3\,h to about 3\,days. Given the fact that there is no known mechanism to accelerate rotation on the
main-sequence, instead, we expect rotational braking and thus longer periods in older clusters, this again seems consistent
with the period range in Praesepe. A more detailed analysis of the rotational evolution based on a comparison of Praesepe
with younger clusters will be given in Sect. \ref{rotage}. Here we conclude that the small sample in Praesepe is
likely to represent a typical range of VLM rotation periods in this cluster.

\subsection{Rotation vs. mass}
\label{rotmass}

It is well-established that rotation is a function of object mass \citep[see][for a review]{2007prpl.conf..297H}. 
Specifically, there is accumulating evidence that in the very low mass regime the average period drops steadily with 
decreasing mass. This positive period-mass correlation at masses $<0.3-0.4\,M_{\odot}$ has been found in the ONC 
\citep{2001ApJ...554L.197H}, $\epsilon$\,Ori \citep{2005A&A...429.1007S}, and the 
Pleiades \citep{2004A&A...421..259S}, clusters with ages between 1 and 125\,Myr. Similarly, in the clusters NGC2516 and 
M34 \citep{2006MNRAS.370..954I,2007MNRAS.tmp..276I}, which roughly mark the ZAMS for very low mass objects at ages of 
150 and 200\,Myr, there seems to be a general decline of the upper envelope of the periods with decreasing mass 
\citep[][see their Fig. 17]{2007MNRAS.tmp..276I}. It has been pointed out that this trend is consistent 
with constant angular momentum for all object masses \citep[e.g.][]{2001ApJ...554L.197H}, and thus might just reflect 
the drop in stellar radius. Since it is already established at very young ages, it may be related to the initial
distribution of angular momentum.

Here we probe if our small period sample in Praesepe -- older than all other clusters with rotation periods in this
mass range -- allows us to see a similar trend. We derived masses for the five objects in Table \ref{periods} by comparing
the available photometry in the I-band from the literature with evolutionary tracks from \citet{1998A&A...337..403B}
for an age of 0.75\,Gyr. According to these estimates, all five objects are in the very low mass regime with masses
$\le 0.4\,M_{\odot}$. Due to age uncertainties, photometric band inconsistencies, and model shortcomings, the
uncertainties in the derived masses are considerable (probably on the order of 50\%). However, most of this uncertainty
is systematic, thus we expect that in a relative sense our masses are realistic. We caution against comparing these
values with masses derived using a different approach. Using the same model isochrone, we determined effective temperatures
for our five targets. Comparing with the $T_{\mathrm{eff}}$ scale from \citet{2003ApJ...593.1093L} gives an estimate
for the spectral types (see Table \ref{periods}). The uncertainties in these 'photometric' spectral types are probably 
$\pm 1-2$ subclasses.

\begin{figure}
\includegraphics[width=6cm,angle=-90]{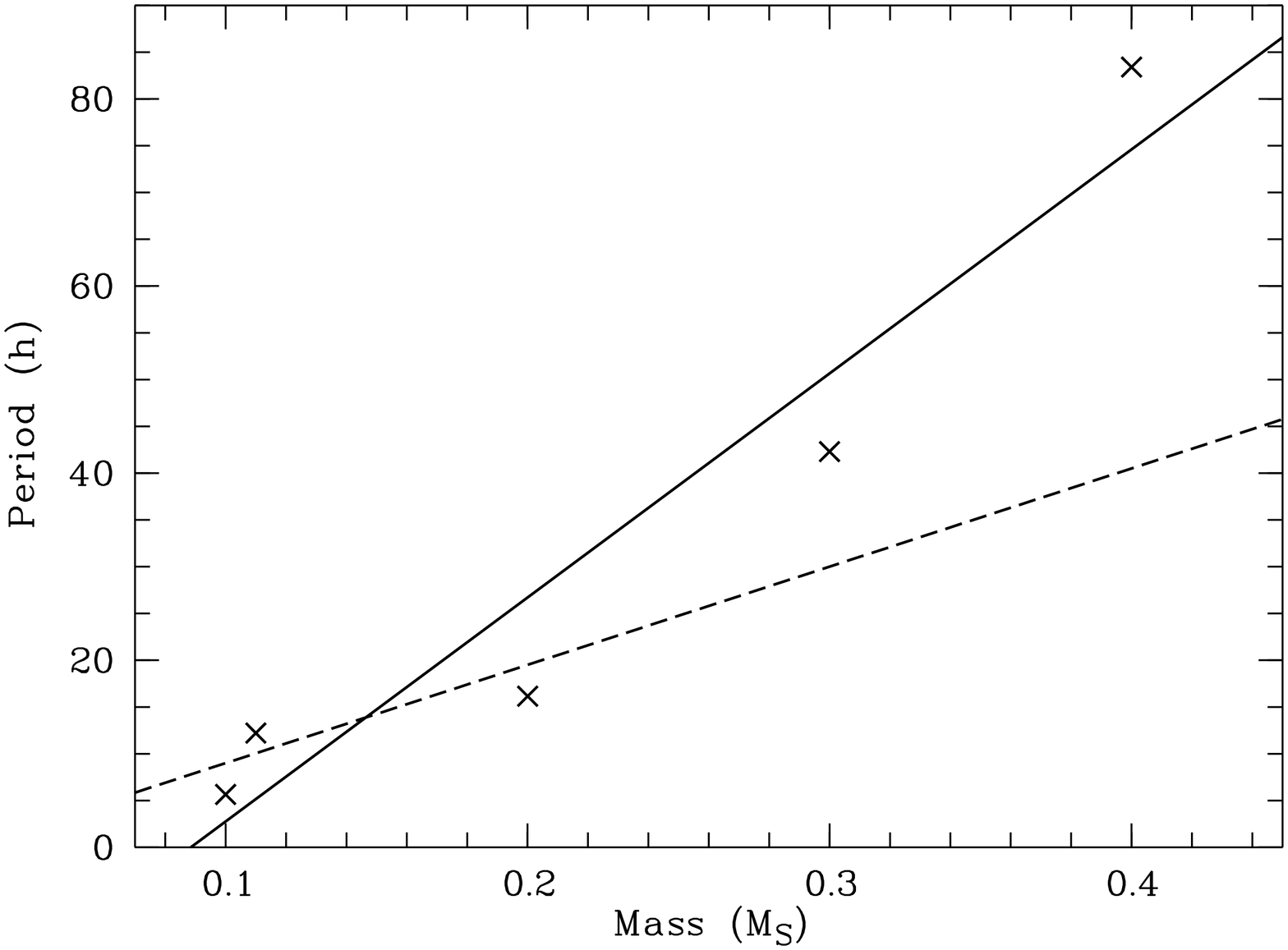} 
\includegraphics[width=6cm,angle=-90]{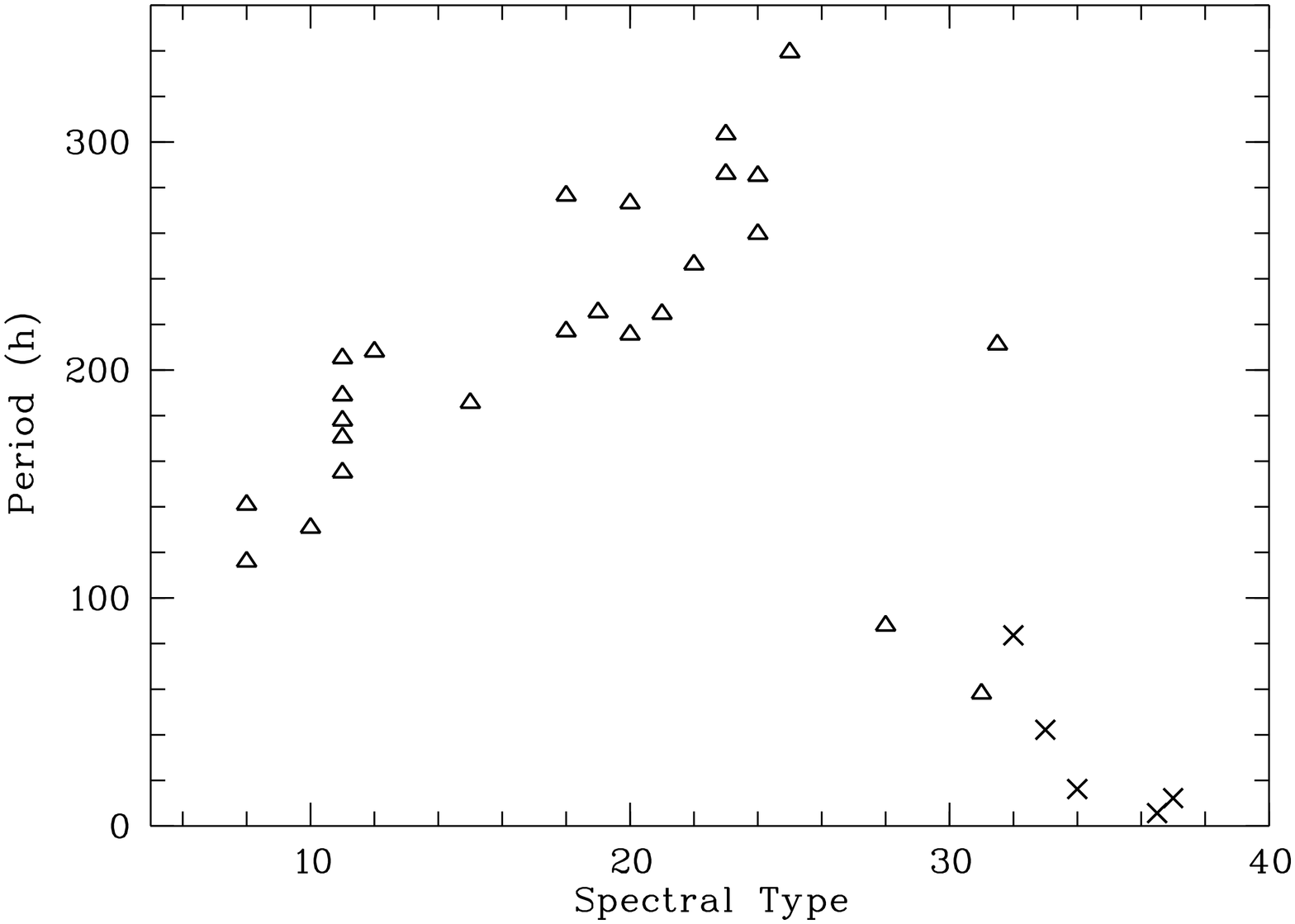} 
\caption{{\bf Upper panel:} Rotation period vs. mass for VLM objects in Praesepe. The solid line is a linear fit; the 
dashed line shows the P(M) fit for our VLM period sample in the Pleiades \citep{2004A&A...421..259S}.
{\bf Lower panel:} Rotation period vs. spectral type (as indicator for mass): Crosses are the five VLM periods in Praesepe 
from this work, triangles are periods in the coeval Hyades from \citet{1987ApJ...321..459R} and \citet{1995PASP..107..211P}. 
Spectral type is parameterised as follows: G0 -- 10, K0 -- 20, M0 -- 30. \label{f2}}
\end{figure}

In Fig. \ref{f2}, upper panel, we plot periods vs. masses for the five objects in Table \ref{periods}. It is immediately
obvious that the periods appear to increase with mass in a roughly linear way. A linear least-square fit (shown as
dotted line) gives: $P = (240 \pm 40)\,(M/M_{\odot}) - (21\pm10)\,h$. The correlation is significant with a false alarm 
probability of 1.0\%. The slope in the relationship is clearly steeper than in the Pleiades ($105 \pm 61\,(M/M_{\odot})$,
\citet{2004A&A...421..259S}), maybe indicating that rotational braking on the main-sequence is a function of object
mass. Based on only five datapoints, however, this finding is of somewhat limited value. Clearly, more datapoints are needed
to solidify the main-sequence P-M correlation in the VLM regime.\footnote{We note that for the three objects with
published H$\alpha$ equivalent widths the rotation periods seem to decline with increasing activity, so it may be
that the observed period-mass trend is in fact a reflection of activity levels increasing with later spectral types.}

It is more instructive to look on the VLM periods in comparison with rotation measurements obtained for more massive stars. 
To our knowledge, such data is not available for Praesepe, but for the Hyades, with 600\,Myr a roughly coeval cluster 
\citep{1981A&A....97..235M,1998A&A...331...81P}. For this comparison, 
we prefer to use spectral types instead of masses, in order not to be biased by model inconsistencies in the mass estimates, 
which are difficult to avoid when covering a mass range of more than one order of magnitude. Since we use only coeval objects, the 
spectral type is a valid indicator for stellar mass. We collected a sample of 25 periods for Hyades members from the photometric 
monitoring campaigns published by \citet{1987ApJ...321..459R} and \citet{1995PASP..107..211P}. Spectral types for these objects 
have originally been published by \cite{1952BAN....11..385V} and \citet{1969AJ.....74....2V}. Combined with our sample, the 
spectral range from late F to late M is covered, roughly corresponding to a mass range from 0.1 to 2\,$M_{\odot}$.

The period/spectral type relation is shown in Fig. \ref{f2}, lower panel. It clearly demonstrates that for F-K stars the
periods increase towards later spectral types. According to \citet{1987ApJ...321..459R}, these periods can be explained in 
terms of the correlation between magnetic activity and the inverse Rossby number $Ro$, the ratio between rotation period and 
convective turnover timescale $\tau_C$ \citep[e.g.][]{1984ApJ...279..763N}. This relation basically implies that the magnetic 
field amplification mainly depends on convection properties and rotation -- supporting the idea of an $\alpha\omega$ dynamo 
operating in F-K stars. We note that the Hyades periods for F-K stars represent what \cite{2003ApJ...586..464B} called
the I-sequence of rotational evolution, i.e. objects with interface dynamo.

As it is apparent from Fig. \ref{f2}, this sequence breaks down roughly at spectral types K8-M2, corresponding 
to a mass of about 0.4-0.6\,$M_{\odot}$ (or a $B-V$ colour $>1.5$). Without clear transition regime, the periods drop
by almost one order of magnitude between K8 and M2. Even if the period distribution at very low masses is not complete
(but see above), the simple fact that the fast rotators do exist among M dwarfs and not in the G-K spectral range
indicates a fundamental change in stellar rotation at the given mass limit.

The underlying physical reason for this breakdown is not clear. Obviously, it indicates a dramatic change in the rotational
braking at a certain mass, which may be related to a change in wind properties, surface magnetic field, or dynamo action. 
It should be noted that the break in rotation period in Praesepe occurs roughly at the same spectral type as the onset of
chromospheric activity measured as H$\alpha$ emission: As recently reported by \citet{2006AJ....132.1517K}, Praesepe
objects earlier than M1 are rarely chromospherically active, in contrast to later type objects. Since enhanced
activity would normally be associated with the potential of efficient angular momentum loss, this poses the question 
why the fast rotating M-dwarfs in Praesepe are the most active objects.

In the rotational evolution scheme by \citet{2003ApJ...586..464B}, the fast rotating M-dwarfs in Praesepe would represent 
the C-sequence (quote: 'possess only a convective field', 'no large scale dynamos'). In this view, the 
mass limit at which I-sequence and C-sequence bifurcate should be a function of age and shift to later spectral types as
objects get older. There is indeed some evidence for this: In $v\sin i$ data for field M dwarfs, most of them certainly 
older than Praesepe, the fraction of fast rotators increases from zero at M3 to 100\% at M6 \citep{1998A&A...331..581D}, 
clearly at lower masses than the transition seen in Praesepe/Hyades. Still, this does not explain {\it why} the magnetic 
field and the dynamo, as suggested by \citet{2003ApJ...586..464B}, should change abruptly at a given mass.

Is there a link between the rotational properties and the interior structure of the stars in the VLM range? Objects
with 0.5\,$M_{\odot}$ are still thought to harbour a substantial radiative core \citep{1997A&A...327.1039C}. Going to 
lower masses, however, the bottom of the convective envelope drops quickly until the objects are fully-convective.
There is strong reason to believe that dynamo action changes in a fundamental way with the disappearance of the radiative 
core. The $\alpha\omega$ dynamo ('interface dynamo') thought to operate in solar-type stars requires a shear layer at the 
bottom of the convection zone, which is not supposed to exist in fully convective objects. Thus, it can be expected that 
the F-K type activity vs. $Ro$ relation breaks down as soon as the radiative core disappears. At the same time, the objects
may lose the ability of efficient rotational braking, possibly due to a change from large-scale to small-scale fields,
resulting in fast rotators.

However, structural models of low-mass stars essentially agree on the fact that this transition is supposed to occur at
masses between 0.3 and 0.4\,$M_{\odot}$ for $M/H = -2 \ldots 0$ \citep[e.g.][]{1997A&A...327.1039C,2000A&A...360..935M} or
even below \citep{2001ApJ...559..353M}, i.e. at lower masses than the observed drop in rotation periods apparent in Fig. 
\ref{f2}. This limit depends on metallicity in a complex way and is expected to increase somewhat for $M/H>0$ 
\citep{1997A&A...327.1039C}. Since both Hyades and Praesepe have higher than solar metallicities ($M/H$ of 0.17 and 
0.14 respectively), this might account for some of the the difference in object mass between observed and expected 
transition to fast rotators. If the break in the mass-period relation is entirely to explain with the change to
fully convective objects, however, remains questionable.

The appearance of Fig. \ref{f2} and its possible potential to probe magnetic and structural properties is intriguing enough 
to justify future research to explain its origin. The following approaches appear to be particularly useful: a) modeling the 
transition to fully convective objects for a range of metallicities, b) complementary observations in the K8-M2 spectral range 
to check for changes in magnetic properties (e.g., Doppler imaging, magnetic field measurements, multi-wavelength monitoring), 
c) rotational studies for stars with masses between 0.6 and 0.3\,$M_{\odot}$, which are so far only represented by 4 datapoints. 

\subsection{Rotation vs. age}
\label{rotage}

The rotational evolution of stars is believed to be controled by mainly three mechanisms:\\ 
a) disk braking, i.e. angular momentum losses due to coupling between star and circumstellar disk, for 
example as so-called 'disk-locking' \citep[see the review by][]{2007prpl.conf..297H},\\ 
b) pre-main sequence contraction resulting in a spin-up,\\
c) magnetic winds driven by dynamo actions, carrying away mass and angular momentum.

\begin{figure}
\includegraphics[width=6cm,angle=-90]{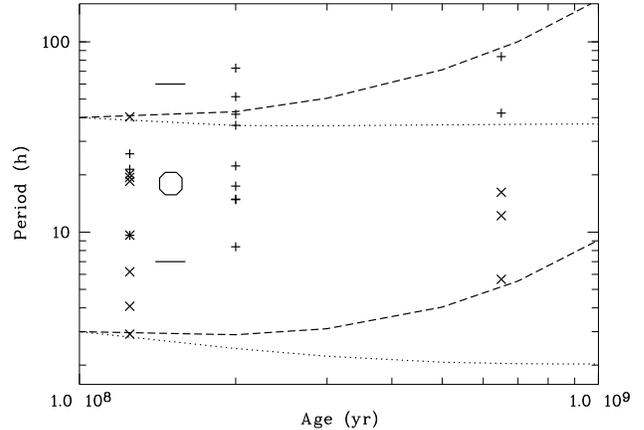} 
\caption{Rotation periods for $0.1-0.4\,M_{\odot}$ mass objects  in Pleiades, NGC2516, M34 
\citep{2004A&A...421..259S,2006MNRAS.370..954I,2007MNRAS.tmp..276I}, and Praesepe (this paper). We chose
different symbols for two mass bins: $0.2-0.4\,M_{\odot}$ -- plus (+), $0.1-0.2\,M_{\odot}$ -- cross (x). For
NGC2516, we plot the median as octagon and the 10\%- and 90\%-percentiles as horizontal lines. Note that the upper
period limits in Pleiades,  M34, and NGC2516 are reliably determined, while the period range in Praesepe may be
incomplete. The dotted lines show the period evolution as expected for zero wind braking. Dashed lines include
exponential wind braking with $\tau = 600$\,Myr. \label{f3}}
\end{figure}

In Fig. \ref{f3}, we compare our rotation periods for Praesepe with period data for VLM objects in younger
clusters: eleven rotation periods for stars in the Pleiades (age 125\,Myr),
\citep{2004A&A...421..259S,1999AJ....118.1814T}, about 250 periods for objects NGC2516 (age 150\,Myr) shown
as median, 10\%, and 90\% percentiles \citep{2007MNRAS.tmp..276I}, nine periods in M34 (age 200\,Myr, 
\citet{2006MNRAS.370..954I}. Together with the five Praesepe periods, this is to our knowledge the total 
available period data for main-sequence VLM stars. In all four clusters, the sampling of the 
photometric monitoring was sensitive to periods ranging from $\le 30$\,min to at least four days.
In the Pleiades and M34, the detection range extends to 12 and $\sim 10$\,days, respectively.  To 
illustrate the differences in the mass ranges, objects with $0.2<M<0.4\,M_{\odot}$ are shown as plus signs 
and lower mass objects as crosses. Note that the majority ($\sim 90$\%) of the periods in NGC2516 has been 
obtained for stars in the upper mass bin. For Praesepe, we use here an age of 650\,Myr, as given by 
\citet{1981A&A....97..235M}; shifting the datapoints within 600-900\,Myr (the likely age range for 
the cluster) does not affect the following results.

We calculated simple rotational evolutionary tracks taking into account spin-up due to contraction and angular 
momentum losses by stellar winds. Since typical disk lifetimes are $<10$\,Myr, the influence of disk braking is 
irrelevant here. The spin-up was estimated using radii from the models of \citet{1997A&A...327.1039C}.
Since even the lowest mass objects have reached their final radii to within 10\% after 200\,Myr, the effect
of contraction of the rotation period is only visible between 100 and 200\,Myr. After that, the rotational evolution 
is almost entirely determined by braking due to stellar winds -- that's why main-sequence clusters are the ideal environment 
to test wind braking. For solar-type stars, wind braking results in the so-called Skumanich law \citep{1972ApJ...171..565S}:
$P \propto \sqrt{t}$. In \citet{2004A&A...421..259S} we demonstrated that this braking law is not applicable in
the pre-main sequence evolution of VLM objects, because it predicts periods $>4$\,days at $\sim 100$\,Myr, while the 
upper VLM period limit at this age, reliably determined from the available period and $v\sin i$ data, is only $\sim 2$ 
days. Thus, a more moderate braking law has to be used, as it is expected for objects with saturated
dynamo \citep{2000AJ....119.1303T,2003ApJ...586..464B} or for objects with a strong concentration of magnetic 
flux near the pole \citep{1997A&A...324..943S}. We use here an exponential form for the braking law 
$P \propto \exp{(t)}$, as it is expected for a saturated dynamo. Given the lack of understanding for the
underlying physics for this type of rotational braking, this should be treated as an ad-hoc solution to 
provide moderate braking, rather than an accurate physical model.

Taken together, the rotational evolution can be expressed as: ($R_i$: initial radius, $R_f$: final radius, $P_i$: initial
period)
\begin{equation}
P_f = \alpha \times (R_i/R_f)^2 \times P_i
\end{equation}
with $\alpha = 1$ for zero wind braking and $\alpha = \exp{(t/\tau)}$ for exponential 
braking, where $\tau$ is the braking (or spin-down) timescale. In Fig. \ref{f3} we plot both cases, no braking with dotted lines, 
exponential braking with dashed lines. The rotational evolution is shown for 
0.4$\,M_{\odot}$, starting at $P_i=40$\,h (upper limit in the Pleiades) and 0.1$\,M_{\odot}$, starting $P_i=3$\,h (lower 
limit in the Pleiades). The different values for $P_i$ were chosen to reflect the P-M relationship discussed in Sect. 
\ref{rotmass}. As can be seen from the tracks, changing the mass does not significantly affect model tracks.

In the discussion of this plot, it is important not to be misguided by the small number of periods. In particular, it is 
essential to keep in mind that the Praesepe period range is probably incomplete (in contrast to the younger clusters). 
Thus, rather than reproducing upper and lower period limits, the rotational evolutionary tracks should simply be able 
to explain the {\it existence} of the periods measured in Praesepe. 

As can be seen in Fig. \ref{f3}, tracks without any braking on the main sequence (dotted lines) have problems reproducing the 
longest periods measured so far in NGC2516, M34, and Praesepe. Thus, some rotational braking on the main-sequence is likely 
occuring among VLM objects, in agreement with our findings in \citet{2004A&A...421..259S}. From the tracks for exponential braking, 
we can rule out that the braking timescale $\tau$ is shorter than 200\,Myr: Such low values for $\tau$ would imply that an 
object with 3\,h rotation period at 100\,Myr (lower limit in the Pleiades) has a period longer than 100\,h at the age of Praesepe 
-- thus not allowing for any of our five measured periods. The plausible range for the braking timescales is between 400\,Myr and 
800\,Gyr (in Fig. \ref{f3} we show the tracks for $\tau = 600$\,Myr). 
\footnote{Based on the currently available data, it appears that the period ranges in M34 and Praesepe are indistinguishable. 
If this is indeed the case, most of the rotational braking in the VLM range would occur at ages $<200$\,Myr. The spindown would 
be more rapid until about the age of M34 and would slow down considerably after that. While such a time-dependent spindown is 
certainly an interesting scenario, we do not put too much trust in this interpretation, because low number statistics and period 
incompleteness in Praesepe might easily be responsible for the effect. In addition, the mass ranges for the objects with periods 
differ slightly from cluster to cluster, which in combination with the strong period-mass relation (see Sect. \ref{rotmass}) can 
lead to biases.}

These results are mostly consistent with earlier findings for the spin-down timescale in the VLM regime. In \citet{2004A&A...421..259S} 
we conclude that the rotational braking in the VLM regime occurs on timescales of a 'few hundred Myrs'. Consistently, 
\citet{2000AJ....119.1303T} find a value of $\tau = 246 \pm 55$\,Myr from a comparison of $v\sin i$ in Pleiades and Hyades.
Several studies point to spin-down timescales in the range of or larger than 1\,Gyr 
\citep{1998A&A...331..581D,2000ApJ...534..335S,2003ApJ...586..464B}. All these estimates are in order-of-magnitude agreement 
with our new constraint. Taken together, the available rotational data for VLM objects favours the occurence of weak rotational
braking on the main-sequence, with exponential spin-down timescales larger than 200\,Myr, maybe as long as a few Gyrs.

According to \citet{2003ApJ...586..464B}, the spin-down timescales for fast rotating G-stars in young open clusters are 
$\sim 30$\,Myr. Thus, among VLM objects, the spin-down happens at a much slower rate than for solar-mass stars: the braking
timescales are 10-100 times longer. This results in fast rotating main-sequence objects, as already evidenced by the rotation
vs. mass plot in Sect. \ref{rotmass}. Both findings, the steep drop in rotation periods and the similarly steep increase
in the spin-down timescales, are complementary observational manifestation of a fundamental change in the angular momentum 
regulation in the VLM regime.

\section{Summary and outlook}

Rotation periods have been measured from photometric monitoring for five stars in the open cluster Praesepe 
(age $\sim$750\,Myr), all with masses $<0.5\,M_{\odot}$. Our work demonstrates that it is possible to obtain reliable 
periods for faint objects at the bottom of the main-sequence using wide-field imagers at medium-sized telescopes, 
as long as enough observing time is available to provide a large number of datapoints and thus a high level of redundancy. 
We show that main-sequence open clusters like Praesepe are ideal environments to probe the long-term angular 
momentum evolution and the underlying regulation mechanisms.

The five periods range from 5\,h to almost 3.5\,days and have been confirmed by various independent period search 
procedures as well as plausibility checks. We attribute these photometric variations to cool, magnetically induced 
spots, co-rotating with the objects -- thus the periods likely correspond to the rotation periods. Comparing the 
small sample of periods in Praesepe with $v\sin i$ data for the coeval Hyades and with periods measured in younger
clusters, we conclude that they give a reasonable first glance on the very low mass (VLM) period distribution in 
Praesepe. Particularly, it is unlikely that Praesepe harbours a large, undetected population of VLM objects with 
periods significantly longer than 4\,days. Still, the big caveat in the discussion of the periods is the small 
number of datapoints. As long as we do not have a more substantial dataset, we refrain from carrying out a vigorous 
statistical analysis and rely instead on more qualitative assessments. 

We find a trend of decreasing period with decreasing mass in the VLM regime, confirming previous claims in
younger clusters. This trend is roughly linear in our small sample with  $P \approx 240\,(M/M_{\odot})\,h$.
To probe the period-mass relation over a broad range of stellar masses, we combine our dataset with periods
in the coeval Hyades from the literature. We find a dramatic change in the periods at spectral type
K8-M2, corresponding to masses of 0.4-0.6$\,M_{\odot}$: Without clear transition regime, the periods drop
from $\sim 10$\,days in mid K-stars to $<4$\,days in early M-stars. Even considering that the 
VLM period sample may be incomplete, the mere existence of the fast rotating M-dwarfs in Praesepe 
indicates a significant change in the rotational regulation. We note that this change coincides
with the onset of chromospheric activity occuring at spectral type M1 in Praesepe \citep{2006AJ....132.1517K}.

Comparing the periods in Praesepe with rotational data in younger clusters (ages of 100-200\,Myr), we find that
some angular momentum loss is likely to occur in the VLM range. The exponential timescale of spin-down due to 
stellar winds is clearly longer than 200\,Myr, most likely between 400 and 800\,Myr, and thus 10-100 times longer 
than in solar-mass stars. Hence, rotational braking on the main-sequence is clearly less efficient in the VLM 
regime, resulting in fast rotators.

Thus, the periods in Praesepe provide evidence for two observational manifestations of a fundamental change in
the rotational regulation: a clear drop of the rotation periods at 0.4-0.6$\,M_{\odot}$ and very long spin-down 
timescales in the VLM regime. This is in line with previous findings: \citet{2000ApJ...534..335S} made an attempt
to model the rotational evolution of VLM objects and conclude that it is impossible to 'simultaneously reproduce 
the observed stellar spin-down in the $0.6-1.1\,M_{\odot}$ range and for stars between 0.1 and 0.6$\,M_{\odot}$',
implying that the solar-type rotational models are not applicable at very low masses. In the evolutionary scheme 
proposed by \citet{2003ApJ...586..464B}, the break between 'I- and C-sequence' (which is identical to the 
break in the period-mass relation reported in this work) can be seen as another way to describe a transition between two 
different rotational regulation regimes. In summary, the solar-type rotational evolution picture of a transition 
dynamo driving a magnetic wind, which leads to Skumanich-type angular momentum losses, does not appear to be applicable 
in the VLM regime.

The underlying physical reason for this fundamental breakdown of the solar-type rotational braking is not clear. 
We argue that it might be possible to explain it with the disappearance of the radiative core and the resulting 
inability to operate a solar-type interface dynamo. However, this transition is thought to occur at masses 
0.3-0.4$\,M_{\odot}$ in solar-metallicity stars, i.e at somewhat lower masses than the drop in rotation periods. 
The metallicity difference between the Sun and young clusters like Praesepe might explain parts of this discrepancy. 
Fundamental changes in the magnetic field structure, the wind properties, or dynamo action (possibly independent
of interior structure) have to be considered as alternative ways to understand the observational findings.

The obvious signature of a fundamental change seen in the rotation periods holds great potential to use rotation
as a probe for magnetic properties and/or interior structure. Follow-up studies aimed a) to enlarge the rotational 
database for VLM objects on the main-sequence and b) to provide complementary activity-related data in this mass 
regime are thus highly desirable to clarify the open questions.

\section*{Acknowledgments}

We thank the referee for a constructive review. AS would like to thank Andrew Collier Cameron, Volkmar Holzwarth, 
and Jerome Bouvier for instructive discussions regarding topics related to this paper. This study benefited from 
unbureaucratic scheduling of observing time by Ulli Thiele, Astronomy Department Head at Calar Alto Observatory, 
and Artie Hatzes, Director at the TLS Tautenburg. We gratefully acknowledge the efforts of the observers 
at the TLS Tautenburg (Christian H{\"o}gner, Uwe Laux, Frank Ludwig) and Calar Alto Observatory (Ana 
Guijarro, Nicolas Cardiel, Manuel Alises, Jesus Aceituno) to provide high-quality data even under 
often not perfect conditions. This work was partially supported by {\it Deutsche Forschungsgemeinschaft} 
(DFG) grant Ei\,409/11-1 and 11-2.
 
\newcommand\aj{AJ} 
\newcommand\araa{ARA\&A} 
\newcommand\apj{ApJ} 
\newcommand\apjl{ApJ} 
\newcommand\apjs{ApJS} 
\newcommand\aap{A\&A} 
\newcommand\aapr{A\&A~Rev.} 
\newcommand\aaps{A\&AS} 
\newcommand\mnras{MNRAS} 
\newcommand\pasa{PASA} 
\newcommand\pasp{PASP} 
\newcommand\pasj{PASJ} 
\newcommand\solphys{Sol.~Phys.} 
\newcommand\nat{Nature} 
\newcommand\bain{Bulletin of the Astronomical Institutes of the Netherlands}

\bibliographystyle{mn2e}
\bibliography{aleksbib}

\begin{thebibliography}{42}
\expandafter\ifx\csname natexlab\endcsname\relax\def\natexlab#1{#1}\fi

\bibitem[{{Adams} {et~al.}(2002){Adams}, {Stauffer}, {Skrutskie}, {Monet},
  {Portegies Zwart}, {Janes}, \& {Beichman}}]{2002AJ....124.1570A}
{Adams} J.~D., {Stauffer} J.~R., {Skrutskie} M.~F., {Monet} D.~G., {Portegies
  Zwart} S.~F., {Janes} K.~A., {Beichman} C.~A., 2002, \aj, 124, 1570

\bibitem[{{Baraffe} {et~al.}(1998){Baraffe}, {Chabrier}, {Allard}, \&
  {Hauschildt}}]{1998A&A...337..403B}
{Baraffe} I., {Chabrier} G., {Allard} F., {Hauschildt} P.~H., 1998, \aap, 337,
  403

\bibitem[{{Barnes}(2001)}]{2001ApJ...561.1095B}
{Barnes} S.~A., 2001, \apj, 561, 1095

\bibitem[{{Barnes}(2003)}]{2003ApJ...586..464B}
---, 2003, \apj, 586, 464

\bibitem[{{Barrado y Navascu{\'e}s} {et~al.}(1998){Barrado y Navascu{\'e}s},
  {Stauffer}, \& {Randich}}]{1998ApJ...506..347B}
{Barrado y Navascu{\'e}s} D., {Stauffer} J.~R., {Randich} S., 1998, \apj, 506,
  347

\bibitem[{{Chabrier} \& {Baraffe}(1997)}]{1997A&A...327.1039C}
{Chabrier} G., {Baraffe} I., 1997, \aap, 327, 1039

\bibitem[{{Delfosse} {et~al.}(1998){Delfosse}, {Forveille}, {Perrier}, \&
  {Mayor}}]{1998A&A...331..581D}
{Delfosse} X., {Forveille} T., {Perrier} C., {Mayor} M., 1998, \aap, 331, 581

\bibitem[{{Fischer} \& {Marcy}(1992)}]{1992ApJ...396..178F}
{Fischer} D.~A., {Marcy} G.~W., 1992, \apj, 396, 178

\bibitem[{{Hambly} {et~al.}(1995{\natexlab{a}}){Hambly}, {Steele}, {Hawkins},
  \& {Jameson}}]{1995MNRAS.273..505H}
{Hambly} N.~C., {Steele} I.~A., {Hawkins} M.~R.~S., {Jameson} R.~F.,
  1995{\natexlab{a}}, \mnras, 273, 505

\bibitem[{{Hambly} {et~al.}(1995{\natexlab{b}}){Hambly}, {Steele}, {Hawkins},
  \& {Jameson}}]{1995A&AS..109...29H}
---, 1995{\natexlab{b}}, \aaps, 109, 29

\bibitem[{{Herbst} {et~al.}(2001){Herbst}, {Bailer-Jones}, \&
  {Mundt}}]{2001ApJ...554L.197H}
{Herbst} W., {Bailer-Jones} C.~A.~L., {Mundt} R., 2001, \apjl, 554, L197

\bibitem[{{Herbst} {et~al.}(2007){Herbst}, {Eisl{\"o}ffel}, {Mundt}, \&
  {Scholz}}]{2007prpl.conf..297H}
{Herbst} W., {Eisl{\"o}ffel} J., {Mundt} R., {Scholz} A., 2007, Protostars and
  Planets V, 297

\bibitem[{{Hodgkin} {et~al.}(1999){Hodgkin}, {Pinfield}, {Jameson}, {Steele},
  {Cossburn}, \& {Hambly}}]{1999MNRAS.310...87H}
{Hodgkin} S.~T., {Pinfield} D.~J., {Jameson} R.~F., {Steele} I.~A., {Cossburn}
  M.~R., {Hambly} N.~C., 1999, \mnras, 310, 87

\bibitem[{{Horne} \& {Baliunas}(1986)}]{1986ApJ...302..757H}
{Horne} J.~H., {Baliunas} S.~L., 1986, \apj, 302, 757

\bibitem[{{Irwin} {et~al.}(2006){Irwin}, {Aigrain}, {Hodgkin}, {Irwin},
  {Bouvier}, {Clarke}, {Hebb}, \& {Moraux}}]{2006MNRAS.370..954I}
{Irwin} J., {Aigrain} S., {Hodgkin} S., {Irwin} M., {Bouvier} J., {Clarke} C.,
  {Hebb} L., {Moraux} E., 2006, \mnras, 370, 954

\bibitem[{{Irwin} {et~al.}(2007){Irwin}, {Hodgkin}, {Aigrain}, {Hebb},
  {Bouvier}, {Clarke}, {Moraux}, \& {Bramich}}]{2007MNRAS.tmp..276I}
{Irwin} J., {Hodgkin} S., {Aigrain} S., {Hebb} L., {Bouvier} J., {Clarke} C.,
  {Moraux} E., {Bramich} D.~M., 2007, \mnras, 276

\bibitem[{{Kafka} \& {Honeycutt}(2006)}]{2006AJ....132.1517K}
{Kafka} S., {Honeycutt} R.~K., 2006, \aj, 132, 1517

\bibitem[{{Luhman} {et~al.}(2003){Luhman}, {Stauffer}, {Muench}, {Rieke},
  {Lada}, {Bouvier}, \& {Lada}}]{2003ApJ...593.1093L}
{Luhman} K.~L., {Stauffer} J.~R., {Muench} A.~A., {Rieke} G.~H., {Lada} E.~A.,
  {Bouvier} J., {Lada} C.~J., 2003, \apj, 593, 1093

\bibitem[{{Magazzu} {et~al.}(1998){Magazzu}, {Rebolo}, {Zapatero Osorio},
  {Martin}, \& {Hodgkin}}]{1998ApJ...497L..47M}
{Magazzu} A., {Rebolo} R., {Zapatero Osorio} M.~R., {Martin} E.~L., {Hodgkin}
  S.~T., 1998, \apjl, 497, L47+

\bibitem[{{Mermilliod}(1981)}]{1981A&A....97..235M}
{Mermilliod} J.~C., 1981, \aap, 97, 235

\bibitem[{{Montalb{\'a}n} {et~al.}(2000){Montalb{\'a}n}, {D'Antona}, \&
  {Mazzitelli}}]{2000A&A...360..935M}
{Montalb{\'a}n} J., {D'Antona} F., {Mazzitelli} I., 2000, \aap, 360, 935

\bibitem[{{Mullan} \& {MacDonald}(2001)}]{2001ApJ...559..353M}
{Mullan} D.~J., {MacDonald} J., 2001, \apj, 559, 353

\bibitem[{{Noyes} {et~al.}(1984){Noyes}, {Hartmann}, {Baliunas}, {Duncan}, \&
  {Vaughan}}]{1984ApJ...279..763N}
{Noyes} R.~W., {Hartmann} L.~W., {Baliunas} S.~L., {Duncan} D.~K., {Vaughan}
  A.~H., 1984, \apj, 279, 763

\bibitem[{{Perryman} {et~al.}(1998){Perryman}, {Brown}, {Lebreton}, {Gomez},
  {Turon}, {de Strobel}, {Mermilliod}, {Robichon}, {Kovalevsky}, \&
  {Crifo}}]{1998A&A...331...81P}
{Perryman} M.~A.~C., {Brown} A.~G.~A., {Lebreton} Y., {Gomez} A., {Turon} C.,
  {de Strobel} G.~C., {Mermilliod} J.~C., {Robichon} N., {Kovalevsky} J.,
  {Crifo} F., 1998, \aap, 331, 81

\bibitem[{{Pinfield} {et~al.}(1997){Pinfield}, {Hodgkin}, {Jameson},
  {Cossburn}, \& {von Hippel}}]{1997MNRAS.287..180P}
{Pinfield} D.~J., {Hodgkin} S.~T., {Jameson} R.~F., {Cossburn} M.~R., {von
  Hippel} T., 1997, \mnras, 287, 180

\bibitem[{{Prosser} {et~al.}(1995){Prosser}, {Shetrone}, {Dasgupta}, {Backman},
  {Laaksonen}, {Baker}, {Marschall}, {Whitney}, {Kuijken}, \&
  {Stauffer}}]{1995PASP..107..211P}
{Prosser} C.~F., {Shetrone} M.~D., {Dasgupta} A., {Backman} D.~E., {Laaksonen}
  B.~D., {Baker} S.~W., {Marschall} L.~A., {Whitney} B.~A., {Kuijken} K.,
  {Stauffer} J.~R., 1995, \pasp, 107, 211

\bibitem[{{Radick} {et~al.}(1987){Radick}, {Thompson}, {Lockwood}, {Duncan}, \&
  {Baggett}}]{1987ApJ...321..459R}
{Radick} R.~R., {Thompson} D.~T., {Lockwood} G.~W., {Duncan} D.~K., {Baggett}
  W.~E., 1987, \apj, 321, 459

\bibitem[{{Roberts} {et~al.}(1987){Roberts}, {Lehar}, \&
  {Dreher}}]{1987AJ.....93..968R}
{Roberts} D.~H., {Lehar} J., {Dreher} J.~W., 1987, \aj, 93, 968

\bibitem[{{Robichon} {et~al.}(1999){Robichon}, {Arenou}, {Mermilliod}, \&
  {Turon}}]{1999A&A...345..471R}
{Robichon} N., {Arenou} F., {Mermilliod} J.-C., {Turon} C., 1999, \aap, 345,
  471

\bibitem[{{Scargle}(1982)}]{1982ApJ...263..835S}
{Scargle} J.~D., 1982, \apj, 263, 835

\bibitem[{{Scholz} \&
  {Eisl{\"o}ffel}(2004{\natexlab{a}})}]{2004A&A...419..249S}
{Scholz} A., {Eisl{\"o}ffel} J., 2004{\natexlab{a}}, \aap, 419, 249

\bibitem[{{Scholz} \&
  {Eisl{\"o}ffel}(2004{\natexlab{b}})}]{2004A&A...421..259S}
---, 2004{\natexlab{b}}, \aap, 421, 259

\bibitem[{{Scholz} \& {Eisl{\"o}ffel}(2005)}]{2005A&A...429.1007S}
---, 2005, \aap, 429, 1007

\bibitem[{{Schrijver} \& {Zwaan}(2000)}]{2000ssma.book.....S}
{Schrijver} C.~J., {Zwaan} C., 2000, {Solar and Stellar Magnetic Activity}.
  Solar and stellar magnetic activity / Carolus J.~Schrijver, Cornelius Zwaan.~
  New York : Cambridge University Press, 2000.~(Cambridge astrophysics series ;
  34)

\bibitem[{{Sills} {et~al.}(2000){Sills}, {Pinsonneault}, \&
  {Terndrup}}]{2000ApJ...534..335S}
{Sills} A., {Pinsonneault} M.~H., {Terndrup} D.~M., 2000, \apj, 534, 335

\bibitem[{{Skumanich}(1972)}]{1972ApJ...171..565S}
{Skumanich} A., 1972, \apj, 171, 565

\bibitem[{{Solanki} {et~al.}(1997){Solanki}, {Motamen}, \&
  {Keppens}}]{1997A&A...324..943S}
{Solanki} S.~K., {Motamen} S., {Keppens} R., 1997, \aap, 324, 943

\bibitem[{{Stauffer} {et~al.}(1997){Stauffer}, {Hartmann}, {Prosser},
  {Randich}, {Balachandran}, {Patten}, {Simon}, \&
  {Giampapa}}]{1997ApJ...479..776S}
{Stauffer} J.~R., {Hartmann} L.~W., {Prosser} C.~F., {Randich} S.,
  {Balachandran} S., {Patten} B.~M., {Simon} T., {Giampapa} M., 1997, \apj,
  479, 776

\bibitem[{{Terndrup} {et~al.}(1999){Terndrup}, {Krishnamurthi}, {Pinsonneault},
  \& {Stauffer}}]{1999AJ....118.1814T}
{Terndrup} D.~M., {Krishnamurthi} A., {Pinsonneault} M.~H., {Stauffer} J.~R.,
  1999, \aj, 118, 1814

\bibitem[{{Terndrup} {et~al.}(2000){Terndrup}, {Stauffer}, {Pinsonneault},
  {Sills}, {Yuan}, {Jones}, {Fischer}, \&
  {Krishnamurthi}}]{2000AJ....119.1303T}
{Terndrup} D.~M., {Stauffer} J.~R., {Pinsonneault} M.~H., {Sills} A., {Yuan}
  Y., {Jones} B.~F., {Fischer} D., {Krishnamurthi} A., 2000, \aj, 119, 1303

\bibitem[{{van Altena}(1969)}]{1969AJ.....74....2V}
{van Altena} W.~F., 1969, \aj, 74, 2

\bibitem[{{van Bueren}(1952)}]{1952BAN....11..385V}
{van Bueren} H.~G., 1952, \bain, 11, 385

\end{thebibliography}

\label{lastpage}

\end{document}